\documentclass[aip,apl,reprint,superscriptaddress]{revtex4-1}
\usepackage{graphicx} %
\usepackage{dcolumn} %
\usepackage{bm} %
\usepackage{amsmath}
\usepackage{color}
\draft 

\begin{document}


\title{Coupling of a locally implanted rare-earth ion ensemble to a superconducting micro-resonator.} 


%

\author{I. S. Wisby}
\affiliation{National Physical Laboratory, Hampton Road, Teddington, TW11 0LW, UK}
\affiliation{Royal Holloway, University of London, Egham, TW20 0EX, UK}
\author{S. E. de Graaf}
\affiliation{Department of Microtechnology and Nanoscience, MC2, Chalmers University of Technology, SE-41296 Gothenburg, Sweden}
\author{R. Gwilliam}
\affiliation{Advanced Technology Institute, Faculty of Electronics and Physical Sciences, University of Surrey, Guildford, Surrey, GU2 7XH, UK}
\author{A. Adamyan}
\affiliation{Department of Microtechnology and Nanoscience, MC2, Chalmers University of Technology, SE-41296 Gothenburg, Sweden}
\author{S. E. Kubatkin}
\affiliation{Department of Microtechnology and Nanoscience, MC2, Chalmers University of Technology, SE-41296 Gothenburg, Sweden}
\author{P. J. Meeson}
\affiliation{Royal Holloway, University of London, Egham, TW20 0EX, UK}
\author{ A. Ya.~Tzalenchuk}
\affiliation{National Physical Laboratory, Hampton Road Teddington, TW11 0LW, UK}
\affiliation{Royal Holloway, University of London, Egham, TW20 0EX, UK}
\author{T. Lindstr\"{o}m}
\affiliation{National Physical Laboratory, Hampton Road Teddington, TW11 0LW, UK}

\email[]{ilana.wisby@npl.co.uk}


\date{\today}

\begin{abstract}
We demonstrate the coupling of rare-earth ions locally implanted in a substrate (Gd$^{3+}$ in Al$_{2}$O$_{3}$) to a superconducting NbN lumped-element micro-resonator. The hybrid device is fabricated by a controlled ion implantation of rare-earth ions in well-defined micron-sized areas, aligned to lithographically defined micro-resonators. The technique does not degrade the internal quality factor of the resonators which remain above $10^5$. Using microwave absorption spectroscopy we observe electron-spin resonances in good agreement with numerical modelling and extract corresponding coupling rates of the order of $1$~MHz and spin linewidths of $50 - 65$~MHz.
\end{abstract}


\maketitle 


In recent years, much of the rapid progress in solid state quantum information processing has come from the field of circuit quantum electrodynamics (cQED) where a superconducting qubit is coupled to a superconducting resonator\cite{Wallraff2004}. However, it has been found that performance is often fundamentally limited by decoherence of the qubit\cite{Martinis2005}. This has prompted interest in a hybrid approach, combining superconducting circuits with other two-level systems (TLS) in order to utilize the unique strengths of individual systems in conjunction \cite{RevModPhys.85.623}. Such hybrid devices can therefore meet the requirements of long storage times, fast processing speeds as well as coherent information transfer.

A hybrid system under investigation is the coupling of spin degrees of freedom in natural systems (eg. cold atoms and ions \cite{blinov2004observation, Rosenfeld2007,PhysRevLett.103.043603}, molecules \cite{andre2006coherent,Rabl2006}, two-level defects \cite{neeley2008process, Falk2013} and spin ensembles \cite{Steger2012,PhysRevB.81.241202}) to superconducting microwave resonators. Such systems aim to exploit the long coherence times provided by natural systems largely decoupled from the environment, alongside the fast-processing capabilities of cQED. Spin doped crystals are particularly suitable for quantum memory applications \cite{Imamoglu2009, Wesenberg2009} having recently demonstrated exceptionally long coherence times\cite{Saeedi15112013}. Rare-earth (R.E) ion doped crystals are of particular interest for transducer applications necessary for long-range quantum communications due to both microwave and optical accessibility mediated by their inner $4f$ shell transition \cite{Brien2014}.

 A prerequisite for many applications of hybrid devices is for operation within the strong coupling regime - whereby TLS-cavity coupling interactions $g$ must dominate the dissipation processes of the cavity $\kappa$, and the spin system $\gamma$, therefore $g > \kappa ; \gamma$ - necessary for a coherent and reversible transfer of states. The single-spin coupling rate $g_{c}$ between electromagnetic modes of a superconducting resonator and a magnetic moment of a spin is given by $g_{c} = \mu_{b} \sqrt{\mu_{0}\omega_{r}/(2 \hbar V_{c})}$, where $\mu_{b}$ is the magnetic dipole moment of the spin which can be approximated as the Bohr magneton, $\mu_{0}$ the vacuum permeability, $\omega_{r}$ the cavity frequency and $V_{c}$ the cavity mode volume. Whilst the strong coupling regime is difficult to achieve and yet to be realized with a single spin, the regime can be reached by utilizing an ensemble of $N$ spins in a spin doped crystal - providing an enhancement of collective coupling \cite{PhysRev.93.99}, $g_{coll}= g_{c} \sqrt{N}$.
 
Spin doped crystals have previously demonstrated operation in the strong coupling regime through `flip chip' experiments \cite{Kubo2010, tkalcec2014strong} - mechanically pressing, or gluing a spin doped crystal atop a superconducting cavity, as well as more recently through positioning within a $3$D cavity \cite{Bushev20143D}. Whilst these methods demonstrate the underlying physics necessary for such a hybrid device, the drawbacks include a lack of control of the configuration of coupled spins, an introduction of additional two-level-systems from added interfaces increasing dielectric loss \cite{Burnett2013} and as such an increased decoherence, as well as difficulties in scalability and realizing multiplexed configurations.

In this work, we have implemented an alternative approach, utilizing an ion implantation technique to allow for control of both the location and density of spins without introducing additional dielectric interfaces. The hybrid device is fabricated by a controlled ion implantation of R.E ions in well defined micron-sized areas of a substrate, aligned to lithographically defined micro-resonators: a technique easily scaled up using standard lithographic techniques. Using this technique, we demonstrate coupling of systematically implanted gadolinium (Gd$^{3+}$) in a sapphire (Al$_{2}$O$_{3}$) substrate to the electromagnetic modes of a superconducting NbN resonator which is fabricated atop the ensemble.


The data shown in this letter is obtained using a sample consisting of $7$ frequency multiplexed, inductively coupled resonators, with particular focus on a lumped element (LE) device with resonance frequency $ \omega_{r}/2\pi = 3.352$~GHz. This resonator is fabricated directly above a $100 \times 250$~$\mu$m area of Gd$^{3+}$ ions implanted in the R-cut sapphire substrate - which is used due a low concentration of natural impurities and previously demonstrated low dielectric loss \cite{Burnett2014}. 

$80$~nm Ni alignment markers, able to withstand the high annealing temperatures required in the fabrication process are first evaporated atop the wafer using a resist lift-off mask. A low stress $400$~nm SiN mask provides a stopping barrier for the ion implantation and is next deposited using multi-frequency plasma enhanced chemical vapour deposition. Locally defined windows are created in the mask using photolithography and a CF$_{4}$ reactive ion etch, which allows for precise control of the location of the spin ensemble. 

A Gd$^{3+}$ ensemble of isotope $^{160}$Gd with nuclear spin $I = 0$ and ground state $f^{7}S_{7/2}$ is next implanted. The Gd$^{3+}$ substitutes into the Al sites of symmetry C$_{3}$ of the Al$_{2}O_{3}$ hexagonal crystal lattice \cite{Geschwind1961}. The wafers are implanted using a $2$MV Van der Graff heavy ion accelerator manufactured by HVEE. The implantation is carried out at room temperature with a $7^{\circ}$ tilt to the normal of the R-plane, to a dose of $1 \times 10^{14}$~ions/cm$^{2}$ at an energy of $900$~keV. The instantaneous beam current is limited to $1$~$\mu$A to prevent sample heating with the beam scanned at $1$~kHz frequency in x and y over a $5$~cm $\times$ $5$~cm area to provide a uniform irradiation at $1\%$. 
 
After implantation, the contaminated mask is removed using a buffered oxide etch (BOE) bath. The wafer is then annealed by ramping the temperature at a ramp rate of $10^{\circ}$~C/min and is held at $980^{\circ}$~C for $1$~hour, before it is cooled during approximately $4$~hours. This step seeks to improve the wafer surface quality over the implanted regions, as well as to assist the impurities in reaching the correct lattice sites. Post-anneal, the RMS surface roughness was found using AFM to be $4$~nm, comparable to typical values of pristine sapphire wafers. 

\begin{figure}
\includegraphics[scale = 1]{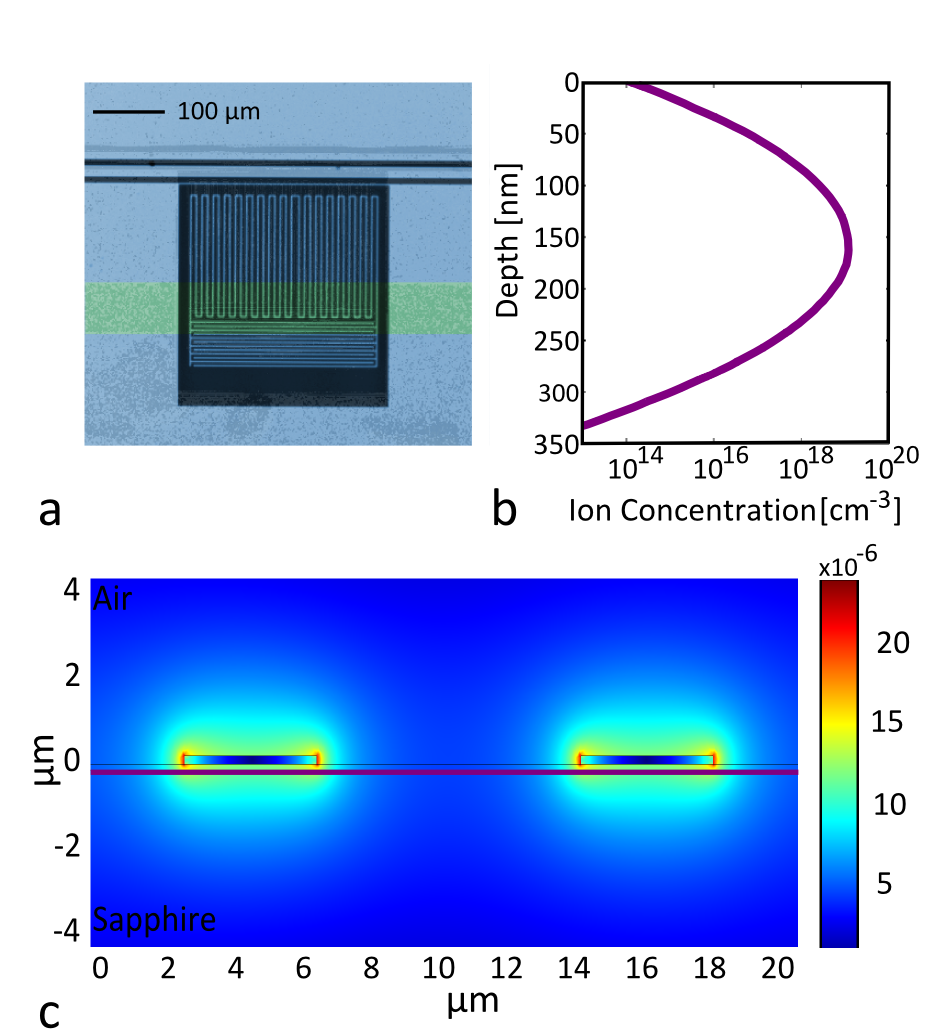}
\caption{\label{Fab} a) Optical image of filtered polarized light showing a LE resonator coupled to the transmission line with the implanted region highlighted (for clarity) by false coloring. b) Concentration profile of ions with respect to depth for an implantation dose $10^{14}$~ions/cm$^{2}$. Peak implantation depth of $160$~nm and FWHM $= 77$~nm.  c) COMSOL$^{\copyright}$ Multiphysics software image of the electromagnetic field of a LE device with the $Gd^{3+}$ implanted region (purple).}
\end{figure}

A $200$~nm NbN sputtered thin film is next deposited, and standard e-beam lithography follows with alignment to the implanted region. An example of a final device is illustrated in Fig. \ref{Fab}a. The implanted region can be observed using an optical confocal microscope with polarization filters and the implanted region is highlighted (for clarity) by false coloring. The area of spins available for coupling to the inductor in this resonator is reduced to a $70 \times 250$~$\mu$m area. 

The resulting implanted spin ensemble distribution is simulated using TRIM$^{\copyright}$ and is a Gaussian profile with peak implantation depth of $160$~nm, a peak concentration of $1.2 \times 10^{19}$~cm$^{-3}$ and a $77$~nm full width at half maximum (FWHM), with the resulting concentration profile shown in Fig. \ref{Fab}b. We calculate a corresponding number of spins available for coupling $N \approx 2.4 \times 10^{11}$. The location of the ensemble with respect to the magnetic field distribution about the inductive meander of our LE device is demonstrated using COMSOL$^{\copyright}$ in Fig. \ref{Fab}c.

The experiment is performed in a dilution refrigerator with base temperature $\approx 20$~mK, equipped with heavily attenuated microwave lines and a low noise cryogenic amplifier. The power in the resonator is approximately $3$~pW. 
 
 Initial characterisation measurements are first performed at base temperature on the $7$ multiplexed resonators through measurement of the microwave transmission coefficient, $S_{21}$, using a vector network analyser (VNA) to obtain phase and magnitude data. Using a numerical fitting function we extract internal quality factors ($Q_{i}$) from this data. All $7$ resonators demonstrated high $Q_{i}$'s, varying between $1-3.5 \times 10^{5}$, dependant on the resonator design - and are comparable to $Q_{i}$'s of reference devices \cite{Lindstrom2009}.
 
\begin{figure}
 \includegraphics[scale = 1]{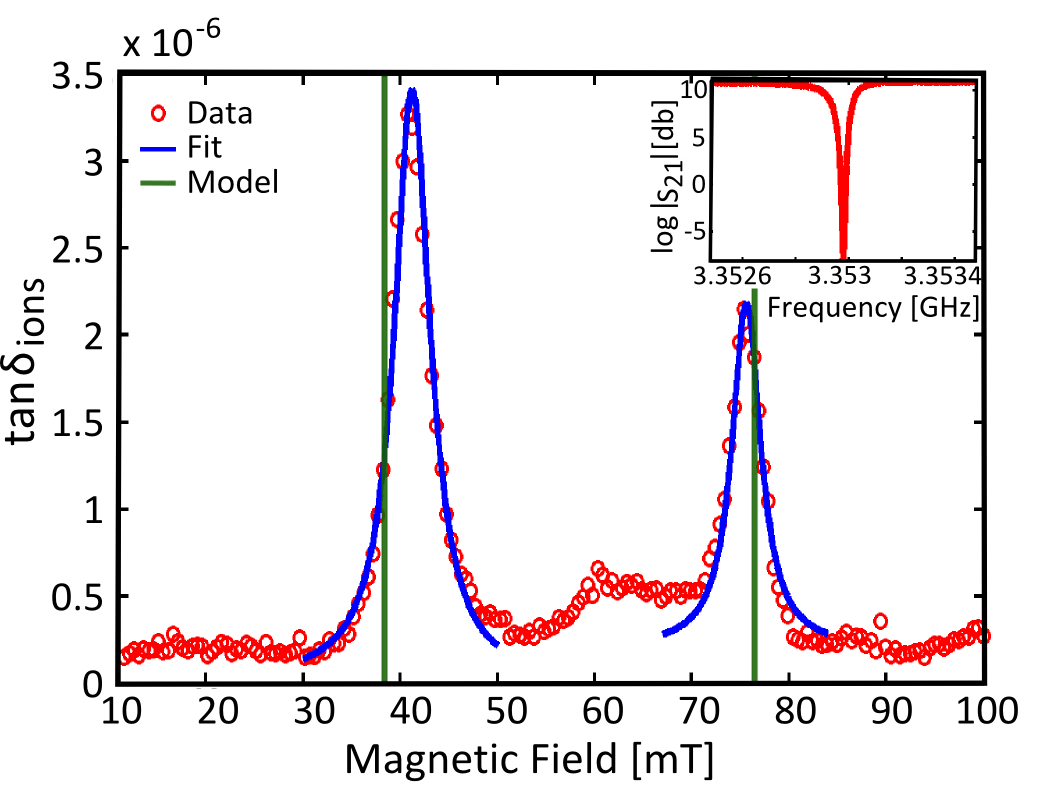}
 \caption{\label{Results} Measurement data $\tan\delta_{ions}$ (red) with spin frequency degenerate at centre frequencies at $ B_{a} = 41$ and $B_{b} = 76$~mT and EASYSPIN$^{\copyright}$ numerical modelling of expected ESR's (green vertical lines) $B_{m_{a}} = 38$ and $B_{m_{b}} = 77$~mT  for comparison.  The spin linewidths, $\gamma_{a,b} = 63, 50$~MHz and collective coupling strengths $g_{coll_{a,b}} = 1.5, 1.0$~MHz, are extracted from the fitting function overlay (blue) using Eq.(\ref{fit}). Inset: Measured $S_{21}$ at $20$~mK, $B=0$~mT. Q$_{i} = 3.3 \times 10^{5}$, Q$_c = 3.8 \times 10^{4}$.}
 \end{figure}

  The $S_{21}$ of the $3.352$~GHz LE resonator presented in this work is found in the inset in Fig. \ref{Results}. The extracted Q$_i = 3.3 \times 10^{5}$ and coupled quality factor Q$_c = 3.8 \times 10^{4}$, gives a resonator dissipation rate of $\kappa = \omega_{r}/Q_{i} = 0.5$~MHz. We next perform absorption spectroscopy on the resonator at $200$~mK. Whilst applying on resonance microwaves ($3.352$~GHz), an external magnetic field $B = 0 - 100$~mT is applied in order to tune the spin ensemble Zeeman transitions into resonance at spin frequency degeneracies.

The $B$ field is applied parallel to the substrate plane, and perpendicular to the microwave propagation, with the in-plane field orientation minimizing flux focusing in the superconductor \cite{Healey2008}. The field is stepped in $\approx 0.2$~mT intervals with wait times to ensure each measurement is in a steady state. At each interval the local minimum is centred on the VNA and $S_{21}$ measured. 
  
We next extract the residual loss tangent due to the ions, $\tan \delta_{\text{ions}} = 1/Q_{\text{ions}}$ from this data: Numerical fitting of the resonator $S_{21}$ response is first used to extract the total measured loss tangent $\tan \delta_{\text{m}} = 1/Q_{\text{m}} = \tan \delta_{\text{c}} + \tan \delta_{\text{int}}$. $\tan \delta_{\text{c}} = 1/Q_{\text{c}}$ is due to coupling to the transmission line, and the intrinsic loss tangent $\tan \delta_{\text{int}}$ can be further subdivided as $\tan \delta_{\text{int}} = \tan \delta_{\text{ions}} + \tan \delta_{\text{diel}} + \tan \delta_{\text{B}}$, corresponding to loss tangents due to the ions, dielectric losses and the external magnetic field, respectively. A polynomial fit is then used to subtract the background $\tan \delta_{\text{diel}} + \tan \delta_{\text{B}}$, providing us $\tan \delta_{\text{ions}}$ alone. The electron-spin resonances (ESR) are therefore observed as an additional absorption mechanism for the microwave photons shown in Fig. \ref{Results} (red). The centre spin frequency degeneracies are found at $B_{a} = 41$ and $B_{b} = 76$~mT. We observe no change in $\omega_{r}$ due to the ions, indicating that we are operating in the weak coupling regime.

 It is also interesting to note that whilst, for example, coplanar Nb resonators display a large degradation in $Q_{i}$ when subject to a $B$ field, we here observe only a $5\%$ degradation when swept to $100$~mT attributable to the use of NbN thin-film, LE geometry and field orientation. It is also interesting to observe that we are sensitive to ESR signals from a much smaller number of spins that would be detectable using standard bulk cavity ESR spectroscopy.
 
In order to understand the features of the ESR spectrum of Gd$^{3+}$:Al$_{2}$O$_{3}$, we use the EASYSPIN$^{\copyright}$ \cite{stoll2006easyspin} software package to model the spin system Hamiltonian:

\begin{equation}
\mathcal{H}=  g \mu_{b} \bf{H} \cdot \bf{S} + H_{\text{ESO}}
\end{equation}

The first term represents the electronic Zeeman splitting with $g = 1.9912$ and the second the high-order extended Stevens operators (ESO) due to the crystal field: $H_{ESO} = \sum B_{\text{k}}^{\text{o}} O_{\text{k}}^{\text{o}}$ where k $= 2,4,6$, o $= 0,3,6$, where each $O_{\text{k}}^{\text{o}}$ is a higher order hermitian spin operator and $B_{\text{k}}^{\text{o}}$ are coefficients parametrized in Ref. \cite{Geschwind1961}. The system has a large zero-field splitting parameter $D = $ $3B_{2}^{0} \approx 3$~GHz. EASYSPIN$^{\copyright}$ numerically diagonalizes this Hamiltonian with respect to our experimental settings, assuming a $B$ field applied parallel to the crystal C-axis. This is adapted for our R-plane cut substrate by a transformation of the reference frame with Euler angles $ \beta = 57.6^{\circ}$, $\alpha = 30^{\circ}$. 

We find our data is in good agreement ($ \approx 95\%$) with numerical modelling, with accessible ESR's indicated as solid vertical lines in Fig. \ref{Results} (green vertical lines): A first order transition is expected at $B_{m_{a}} = 38$~mT and second order transition at and $B_{m_{b}} = 77$~mT when modelled for $\beta = 55^{\circ}$, corresponding to an effective angular error of $- 2 ^{\circ}$. This could be attributable to misalignment in the sample cut, magnetic field alignment errors as well as small deviations in the higher order Stevens operators known from literature \cite{Geschwind1961}.

Features of known impurities within the sapphire: Fe$^{3+}$, Cr$^{3+}$ can be observed at $90$~mT, as well as an unknown signal at $60$~mT, which has previously been observed in Ref.\cite{Farr2013}.

We next model the spin ensemble and cavity as a single mode harmonic oscillator, as performed by Schuster et al. \cite{Schuster2010}. The model is valid so long as $\Delta$, $\gamma$ or $\kappa$ is larger than $g_{coll}$, such that the $Q_{m}$ of a cavity coupled to a spin ensemble is given by
\begin{equation}\label{fit}
Q_{m} = \frac{\Delta^{2} + \gamma^{2}}{2g^{2}_{coll} \gamma + \kappa(\Delta^{2} + \gamma^{2})}\omega_{r}
\end{equation}
with $\Delta$ the detuning from $B_{a,b}$ and the cavity linewidth $\kappa_{a,b}$ taken as $\omega_{r}/Q_{m_{a,b}} = 0.86, 0.47$~MHz respectively. The spin linewidths, $\gamma_{a,b} = 63, 50$~MHz and collective coupling strengths $g_{coll_{a,b}} = 1.5, 1.0$~MHz, are extracted from the fit using Eq.(\ref{fit}).

Numerical modelling has provided approximations of the expected collective coupling $g_{coll_{ex}}$. These were obtained by considering an integration over the magnetic field and implanted ion concentration distributions (see Fig.\ref{Fab}) in calculations of $g_{coll}$. An approximate value for the first order transition $B_{a}$ at $20$~mK provides $ g_{coll_{ex}} \approx 2.8$~MHz and is in reasonable agreement with the extracted value, indicating a good level of control of the implanted spin system.

We are not yet operating in the strong coupling regime -  limited by large $\gamma$ potentially due to inhomogeneous broadening caused by excess spin-spin interactions as well as defects. Previous experiments in Al$_2$O$_{3}$ with $100$~ppm doped $Gd^{3+}$ have reported linewidths down to $22$~MHz \cite{DurhamThesis}, comparable to that of other potential hybrid systems \cite{tkalcec2014strong}. It is therefore speculated that a decrease in $\gamma$ could be achieved with a lesser concentration (whilst maintaining $N$) and reduction in defects, or through use of a different ion/substrate combination. The strong coupling regime could also be reached by further optimization of the collective system by increasing the number of coupled spins. The single spin coupling rate $g_{c}$ can also be maximized further by, for example, operating at a higher centre frequencies. 
 

In conclusion, we demonstrate the coupling of rare-earth Gd$^{3+}$ ions locally implanted in a Al$_{2}$O$_{3}$ substrate to a superconducting NbN lumped element micro-resonator. The hybrid device is fabricated using a  technique for controlled ion implantation of rare-earth's in well-defined micron-sized areas, alongside lithographically defined micro-resonators. Using microwave absorption spectroscopy, we show the collective enhancement of a spin ensemble created via this ion-implantation process as a proof-of-principle of a promising hybrid device. Our technique allows for precise control of the spin ensemble in terms size and location without degradation of $Q_{i}$, as well as scalable integration with lithography defined circuitry and frequency multiplexing technology. We observe electron-spin resonances in good agreement with numerical modelling, corresponding to a collective coupling of the order of $1$~MHz and linewidths of $\approx 50 - 65$~MHz. Whilst the measured collective coupling strengths exceed the decay rate of the cavity, the strong coupling regime is not yet reached due to large spin linewidths. Provided a reduction in linewidths, the presented experiment shows the promising potential of locally implanted rare-earth doped crystals for application in hybrid quantum technologies.

We thank J. Burnett, A.V. Danilov, D. Cox, N. Panjwani and J. Morton for fruitful discussions and support. This work was supported by the NMS, EPSRC, Swedish Research Council (VR) and the Linneqs center. Access to the IBC was supported by the EC program SPIRIT, contract $227012$. 



\bibliography{LS_Library2}

\end{document}